# Run-up characteristics of symmetrical solitary tsunami waves of "unknown" shapes

Ira Didenkulova[1,2], Efim Pelinovsky[1] and Tarmo Soomere[2]

[1] Institute of Applied Physics, Nizhny Novgorod, Russia
[2] Institute of Cybernetics, Tallinn, Estonia

The problem of tsunami wave run-up on a beach is discussed in the framework of the rigorous solutions of the nonlinear shallow-water theory. We present an analysis of the run-up characteristics for various shapes of the incoming symmetrical solitary tsunami waves. It will be demonstrated that the extreme (maximal) wave characteristics on a beach (run-up and draw-down heights, run-up and draw-down velocities and breaking parameter) are weakly dependent on the shape of incident wave if the definition of the "significant" wave length determined on the 2/3 level of the maximum height is used. The universal analytical expressions for the extreme wave characteristics are derived for the run-up of the solitary pulses. They can be directly applicable for tsunami warning because in many case the shape of the incident tsunami wave is unknown.

# 1. Introduction

The reliable estimation of inundation extend is a key problem of coastal wave dynamics, and in particular for tsunami mitigation. Since the characteristic length of a tsunami wave in the coastal zone is several kilometers, the nonlinear shallow water theory is an appropriate theoretical model to describe the process of tsunami run-up on a beach. The problem of the run-up of long non-breaking waves on a plane beach is well described mathematically within the framework of a nonlinear shallow water theory. This approach leads to an analytical solution based on the Carrier–Greenspan transform (Carrier and Greenspan, 1958). Various shapes of the periodic incident wave trains such as the sine wave (Kaistrenko et al. 1991; Madsen and Fuhrman, 2008), cnoidal wave (Synolakis 1991) and nonlinear deformed periodic wave (Didenkulova et al, 2006, 2007b) have been analyzed in the literature. The relevant analysis has also been performed for a variety of solitary waves and single pulses such as soliton (Pedersen and Gjevik 1983; Synolakis 1987; Kânoğlu, 2004), sine pulse (Mazova et al. 1991), Lorentz pulse (Pelinovsky and Mazova 1992), Gaussian pulse (Carrier et al. 2003; Kânoğlu and Synolakis, 2006), *N*-waves (Tadepalli and Synolakis 1994), "characterized tsunami waves" (Tinti and Tonini, 2005) and the random set of solitons (Brochini and Gentile, 2001). However, as is often the case in nonlinear problems, reaching an analytical solution is seldom possible. Run-up of solitary pulses is easily implemented experimentally in measuring flumes, and various experimental expressions are available; see (Madsen and Fuhrman, 2008) for references.

The existing results for the water wave field are based on various initial conditions (shapes of the incident waves) and are therefore not directly comparable with each other. Sometimes, the shape of the incident wave is unknown, and this situation is typical for tsunamis. To get universal expressions for run-up characteristics several parameters can be used. Madsen and Fuhrman (2008) suggest expressing of the formula for run-up height in terms of a surf-similarity. These expressions are applicable for non-breaking waves on a plane beach and for breaking waves as well. Didenkulova et al (2007a, 2008) parameterize run-up expressions using various definitions of the wavelength of non-breaking wave pulses. Below we demonstrate that the definition of wavelength on the 2/3 level from a maximal value (as the "significant wavelength" in physical oceanography and ocean engineering) is optimal. In this case formulas for various extreme run-up characteristics (run-up and draw-down heights and velocities, breaking parameter) are universal and the influence of the initial wave form on extreme run-up characteristics is weak. This result is obtained for incident symmetrical solitary waves.

The paper is organized as follows. The analytical theory of long wave run-up on a beach in the framework of shallow-water theory is briefly described in section 2. Numerical computations of the tsunami waves far from the beach and on the shoreline are based on spectral Fourier series (section 3). The parameterization of wave shapes in formulas for the extreme (maximal) wave characteristics on a beach are discussed in section 4. The main results are summarized in section 5.

## 2. Analytical theory of the long wave run-up on a beach

The run-up of tsunami waves on a beach can be described in the framework of the nonlinear shallow-water equations. If the wave propagates perpendicularly to the isobaths, basic equations are

$$\frac{\partial \eta}{\partial t} + \frac{\partial}{\partial x}\left[(h(x)+\eta)u\right] = 0, \tag{1}$$

$$\frac{\partial u}{\partial t} + u\frac{\partial u}{\partial x} + g\frac{\partial \eta}{\partial x} = 0, \tag{2}$$

where $\eta(x,t)$ is the vertical displacement of the water surface, $u(x,t)$ – depth-averaged water flow, $h(x)$ – unperturbed water depth, and $g$ is the gravitational acceleration. Analytical solutions of this system are obtained for a plane beach only, where the depth $h(x) = -\alpha x$ (Fig. 1). The procedure of the solution is based on the hodograph transformation firstly described in pioneering work by Carrier-Greenspan (1958), and reproduced in different papers cited in section 1.

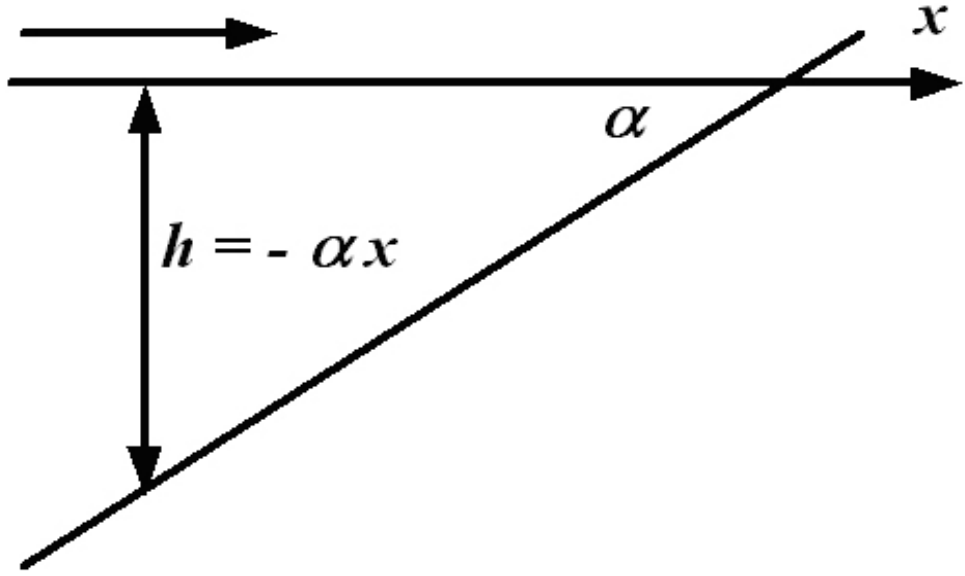

**Fig. 1.** Sketch of the problem

According to this method all variables can be expressed through the "nonlinear" wave function $\Phi(\sigma,\lambda)$ by means of the hodograph transformation:

$$\eta = \frac{1}{2g}\left(\frac{\partial \Phi}{\partial \lambda} - u^2\right), \qquad (3)$$

$$u = \frac{1}{\sigma}\frac{\partial \Phi}{\partial \sigma}, \qquad (4)$$

$$t = \frac{1}{\alpha g}\left(\lambda - \frac{1}{\sigma}\frac{\partial \Phi}{\partial \sigma}\right), \qquad (5)$$

$$x = \frac{1}{2\alpha g}\left(\frac{\partial \Phi}{\partial \lambda} - u^2 - \frac{\sigma^2}{2}\right), \qquad (6)$$

and the wave function, $\Phi(\lambda,\sigma)$ satisfies to the cylindrical linear wave equation

$$\frac{\partial^2 \Phi}{\partial \lambda^2} - \frac{\partial^2 \Phi}{\partial \sigma^2} - \frac{1}{\sigma}\frac{\partial \Phi}{\partial \sigma} = 0. \tag{7}$$

The variables $\lambda$ and $\sigma$ have the meaning of generalized coordinates. Since

$$\sigma = 2\sqrt{g(-\alpha x + \eta)}, \tag{8}$$

the point $\sigma = 0$ corresponds to the instantaneous position of the shoreline (called moving shoreline in what follows).

It is interesting to note that if we analyze the linear system of shallow-water wave theory

$$\frac{\partial \eta}{\partial t} + \frac{\partial}{\partial x}\left[(-\alpha x)u\right] = 0, \tag{9}$$

$$\frac{\partial u}{\partial t} + g\frac{\partial \eta}{\partial x} = 0, \tag{10}$$

the linear version of the hodograph transformation

$$\eta_l = \frac{1}{2g}\left(\frac{\partial \Phi_l}{\partial \lambda_l}\right), \quad u_l = \frac{1}{\sigma_l}\frac{\partial \Phi_l}{\partial \sigma_l}, \quad t = \frac{\lambda_l}{\alpha g}, \quad x = -\frac{\sigma_l^2}{4\alpha g}, \tag{11}$$

transforms (9)-(10) to the wave equation

$$\frac{\partial^2 \Phi_l}{\partial \lambda_l^2} - \frac{\partial^2 \Phi_l}{\partial \sigma_l^2} - \frac{1}{\sigma_l}\frac{\partial \Phi_l}{\partial \sigma_l} = 0, \tag{12}$$

which coincides with the wave equation (7) in the nonlinear problem. However in this case the point $\sigma_l = 0$ corresponds to the unperturbed shoreline ($x = 0$).

Tsunami waves in the deep ocean have small amplitudes and can be described by the linear theory with a very high accuracy. For such an incident wave the boundary conditions for the "nonlinear" (7) and "linear" (12) wave equations coincide, provided they are defined in a far and deep enough area. Consequently, the solutions of the nonlinear and linear problems also coincide in terms of solutions of the wave equation, as the functional forms of its "linear" and "nonlinear" solutions coincide: $\Phi(\sigma,\lambda) \equiv \Phi_l(\sigma_l,\lambda_l)$. Moreover, if the "linear" solution $\Phi_l(\sigma_l,\lambda_l)$ is known, the solution of the nonlinear problem (1)–(2) can be directly found from expressions (3)-(6). In particular, the description of properties of the moving shoreline $\sigma(x,t) = 0$ is straightforward. If the velocity of water particles on the unperturbed shoreline ($x = 0$) is calculated in frames of the linear theory, a real "nonlinear" velocity of the moving shoreline can be expressed in an implicit form (Pelinovsky and Mazova, 1992, Didenkulova et al., 2007b):

$$u(t) = U(\tilde{t}), \text{ where } \tilde{t} = t + \frac{u(t)}{\alpha g}. \tag{13}$$

Mathematically the function $U(t)$ in the linear theory is defined as $U(t) = \lim_{\sigma_l \to 0} u_l$.

The described features allow use of a rigorous "two-step" method to calculate the run-up characteristics if the linear theory describes well their motion far offshore. Firstly, the wave properties on the unperturbed shoreline ($x = 0$) such as the vertical displacement $R(t)$ or the velocity of wave propagation $U(t)$

$$U(t) = \frac{1}{\alpha}\frac{dR}{dt}, \tag{14}$$

are determined from the linear shallow-water theory. Secondly, the properties of the solution of the nonlinear problem (e.g. the real "nonlinear" speed of the moving shoreline) are found from Eq. (13). Finally, the vertical displacement of the water level and position of the shoreline (equivalently, the horizontal extent of the inundation) at any instant of time is

$$r(t) = \eta(t,\sigma = 0) = R\left(t + \frac{u}{\alpha g}\right) - \frac{u^2}{2g}. \tag{15}$$

The important conclusion from Eqs. (13) and (15) is that the maxima of vertical displacements (equivalently, the run-up height or the draw-down depth) and the velocity of the shoreline displacement in the linear and nonlinear theories coincide as noted by Carrier and Greenspan (1958) and Synolakis (1987, 1991) and rigorously demonstrated by Pelinovsky and Mazova (1992).

Another important outcome from Eqs. (13) and (15) is the simple definition of the conditions for the first breaking of waves on a beach. The temporal derivative of the velocity of the moving shoreline, found from Eq. (13), approaches infinity (equivalently, wave breaking occurs) when

$$Br = \frac{\max(dU/dt)}{\alpha g} = 1. \tag{16}$$

This condition has a simple physical interpretation: the wave breaks if the maximal acceleration of the shoreline $R''\alpha^{-1}$ along the sloping beach exceeds the along-beach gravity component $\alpha g$. This interpretation is figurative, because formally $R''$ only presents the vertical acceleration of the shoreline in the linear theory and the "nonlinear" acceleration $du/dt$ (that is not explicitly calculated here) is what actually approached infinity at the moment of breaking.

## 3. Method for computing extreme run-up characteristics

Following to the "two-step" method described above, the linear theory can be used for the computation of extreme characteristics of the tsunami wave run-up. An effective method for solving linear partial differential equations is the Fourier method and its generalizations (for instance, the Hankel transformation for cylindrical wave equation). It is convenient to describe the wave field in terms of its complex (amplitude-phase) spectrum $A(\omega)$ (equivalently, Fourier integral of the associated sea level variations). The particular bounded solution of the cylindrical wave equation can be represented by the Fourier integral

$$\eta(x,t) = \int_{-\infty}^{+\infty} A(\omega) J_0\left(\frac{2\omega|x|}{\sqrt{gh(x)}}\right) \exp(i\omega t) d\omega, \tag{17}$$

where $J_0(y)$ is the zero-order Bessel function. The spectrum $A(\omega)$ can be found from the spectrum $H(\omega)$ of the incident wave with the use of the asymptotic representation of the wave field (17) at $x \to -\infty$ as the superposition of the incident $\eta_+$ and reflected $\eta_-$ waves

$$\eta(x \to -\infty, t) = \eta_+[x, t + \tau(x)] + \eta_-[x, t - \tau(x)]. \tag{18}$$

Here

$$\tau(x) = \frac{2|x|}{\sqrt{gh(x)}} = \int_{-|x|}^{0} \frac{dx}{\sqrt{gh(x)}} \qquad (19)$$

is the travel time from a given location $x$ to the unperturbed original shoreline. This measure can also be interpreted as the phase shift of the reflected wave in space.

In the same limit $x \to -\infty$, Eq. (17) gives

$$\eta_{\pm}(x \to \infty, t) = \frac{1}{\sqrt{2\pi\tau(x)}} \int_{-\infty}^{+\infty} \frac{A(\omega)}{\sqrt{|\omega|}} \exp\left[i\left(\omega(t \pm \tau(x)) \mp \frac{\pi}{4}\operatorname{sign}(\omega)\right)\right] d\omega. \qquad (20)$$

We assume that the incident wave at a fixed point $|x| = L$ (located far away from the shoreline) is specified by the Fourier integral

$$\eta_{+}(t) = \int_{-\infty}^{+\infty} H(\omega) \exp(i\omega t) d\omega. \qquad (21)$$

Its complex spectrum $H(\omega)$ is easily found in an explicit form in terms of the inverse Fourier transform:

$$H(\omega) = \frac{1}{2\pi} \int_{-\infty}^{+\infty} \eta_{+}(t) \exp(-i\omega t) dt. \qquad (22)$$

We assume now that the point $|x| = L$ is located so far from the unperturbed shoreline that decomposition (20) (that formally is correct for $x \to -\infty$) can be used at this point. Comparison of Eqs. (20) and (22) then reveals that

$$A(\omega) = \sqrt{2\pi |\omega| \tau(L)} \exp\left[\frac{i\pi}{4}\operatorname{sign}(\omega)\right] H(\omega). \qquad (23)$$

The solution in Eq. (17) is thus completely determined by the incident wave.

The vertical displacement at the unperturbed shoreline $x = 0$ is a function of the location $L$

$$R(t) = \sqrt{2\pi\tau(L)} \int_{-\infty}^{+\infty} \sqrt{|\omega|} H(\omega) \exp\left\{i\left(\omega(t-\tau(L)) + \frac{\pi}{4}\text{sign}(\omega)\right)\right\} d\omega. \tag{24}$$

The "linear" horizontal velocity of water particles at this point ($x = 0$) can be found from Eq. (14). As was mentioned above, the extreme wave amplitudes (understood as the maximum displacement of water surface) and velocities at the unperturbed shoreline $x = 0$ in linear theory coincide with the maximum run-up (draw-down) heights and velocities in the nonlinear theory. Therefore, we would like to emphasize that solving of nonlinear equations is not necessary if only extreme characteristics of tsunami waves are analyzed.

Integral properties of a wave run-up dynamics also can be found from the linear theory. For instance, an integrated vertical displacement of the shoreline ("set-up") is

$$\hat{R} = \int_{-\infty}^{+\infty} R(t) dt = \sqrt{2\pi\tau(L)} \int_{-\infty}^{+\infty} dt \int_{-\infty}^{+\infty} \sqrt{|\omega|} H(\omega) \exp\left\{i\left(\omega(t-\tau(L)) + \frac{\pi}{4}\text{sign}(\omega)\right)\right\} d\omega. \tag{25}$$

After changing the order of integration and taking into account the properties of delta-function Eq. (25) can be rewritten as

$$\hat{R} = 2\pi\sqrt{2\pi\tau(L)} \int_{-\infty}^{+\infty} \sqrt{|\omega|} H(\omega) \exp[i(\pi/4)\text{sign}(\omega)]\delta(\omega) d\omega. \tag{26}$$

A physical sense of $H(0)$ is an integrated displacement of the incident wave, which is bounded. Since the integrand is a continuous function, the integral in Eq. (26) is equal to zero. Therefore, the tsunami run-up is always presented as reversal oscillations of the shoreline, and a run-up phase changes into a receding phase and this process does not depend on the shape of the incident wave.

Let us consider an incident tsunami wave having pulse shape (for example, a positively defined disturbance – wave of elevation or crest) with amplitude $H_0$ and duration $T_0$ at $|x| = L$ propagating onshore. It can be nondimensionalized as

$$\eta(t) = H_0 f(t/T_0); \qquad f(\zeta) = \int_{-\infty}^{\infty} B(\Omega) \exp(i\Omega\zeta) d\Omega, \tag{27}$$

where

$$\zeta = t/T_0, \qquad \Omega = \omega T_0, \qquad B(\Omega) = \frac{1}{2\pi}\int_{-\infty}^{\infty} f(\zeta)\exp(-i\Omega\zeta)d\zeta. \qquad (28)$$

In this case, formulas for velocity of the moving shoreline (14) and linear acceleration, connected with the breaking parameter of the wave (16), and for the maximal vertical displacement (24), can be presented as

$$R_{max} = R_0 p_R, \qquad p_R = \max\{I\}, \qquad I = \int_{-\infty}^{\infty}\sqrt{|\Omega|}B(\Omega)\exp\left[i\left(\Omega\zeta + \frac{\pi}{4}sign(\Omega)\right)\right]d\Omega, \qquad (29)$$

$$U_{max} = \frac{R_0}{\alpha T_0} p_U, \qquad p_U = \max\left\{\frac{dI}{d\zeta}\right\}, \qquad (30)$$

$$(dU/dt)_{max} = \frac{R_0}{\alpha T_0^2} p_{Br}, \qquad p_{Br} = \max\left\{\frac{d^2I}{d\zeta^2}\right\}, \qquad (31)$$

$$R_0 = \sqrt{\frac{4\pi L}{\lambda_0}}H_0, \qquad \lambda_0 = \sqrt{gh_0}T_0, \qquad (32)$$

where $\lambda_0$ is the wavelength and $h_0$ is the water depth at $|x| = L$.

In many cases it is not clear how to determine the wavelength $\lambda_0$ and the duration $T_0$ of a solitary pulse. In particular, most of the wave shapes (represented by analytical functions that are continuous in all derivatives) are nonzero everywhere at $-\infty < t < \infty$. There is obvious ambiguity in the definition of their wavelength (or duration) that can be interpreted as their width at any level of elevation, or by the value of an appropriate integral (Didenkulova et al, 2007a, 2008).

A convenient definition of the wavelength is the extension (spatial or temporal) of the wave profile elevation exceeding the 2/3 level of the maximum wave height. This choice is inspired by the definition of the significant wave height and length in physical oceanography and ocean engineering. For symmetric solitary waves, "significant" wave duration and "significant" wavelength are

$$T_s = 2T_0 f^{-1}\left(\frac{2}{3}\right), \qquad \lambda_s = \sqrt{gh_0}T_s, \qquad (33)$$

where $f^{-1}$ is the inverse function of $f$. Thus the formulas for the maximal displacement, the velocity of the moving shoreline, and the breaking parameter can be expressed as

$$R_{\max} = \mu_R^+ H_0 \sqrt{\frac{L}{\lambda_s}}, \qquad U_{\max} = \mu_U^+ \frac{H_0 L}{\lambda_s} \sqrt{\frac{g}{\alpha \lambda_s}}, \qquad Br = \mu_{Br} \frac{H_0 L}{\alpha \lambda_s^2} \sqrt{\frac{L}{\lambda_s}}, \qquad (34)$$

where coefficients (below called form factors) $\mu_R^+$, $\mu_U^+$ and $\mu_{Br}$ depend on the wave form:

$$\mu_R^+ = 2\sqrt{2\pi f^{-1}\left(\frac{2}{3}\right)} p_R, \qquad \mu_U^+ = 4\sqrt{2\pi \left[f^{-1}\left(\frac{2}{3}\right)\right]^3} p_U, \qquad \mu_{Br} = 8\sqrt{2\pi \left[f^{-1}\left(\frac{2}{3}\right)\right]^5} p_{Br}. \qquad (35)$$

The analogous formulas for draw-down height and velocity can be obtained from (34, 35) by replacing $p_R \to \bar{p}_R = \min\{I\}$, $p_U \to \bar{p}_U = \min\{dI/d\zeta\}$ and $p_{Br} \to \bar{p}_{Br} = \min\{d^2I/d\zeta^2\}$ in (29) and (30).

A remarkable property of this choice is that if the solitary wave duration is determined at the 2/3 level of the maximum height (33), the effect of the difference in the wave shapes will be fairly small. The analytical expressions for maximal run-up characteristics (run-up and draw-down heights, run-up and draw-down velocities and breaking parameter) become universal and depend on height and duration of the incoming onshore wave only.

## 4. Results of calculations

Let us first consider the run-up of incident symmetrical positive waves having the shape of various "powers" of a sinusoidal pulse

$$f(\zeta) = \cos^n(\pi \zeta), \qquad \text{where } n = 2, 3, 4, \ldots \qquad (36)$$

which are defined on the segment $[-1/2, 1/2]$. Their shapes have a certain similarity, but their wave characteristics, such as mean water displacement, energy and wave duration on various levels, differ considerably (Fig. 2). The functions representing such impulses have different smoothness: their $n$-th order derivatives are discontinuous at their ends. The case of $n = 1$ is not considered, as the relevant integrals in Eqs. (29) and (30) do not converge. The run-up of sinusoidal pulse for the case of $n = 2$ is presented on Fig. 3. Such oscillations are typical for run-up of symmetric solitary waves on a beach of constant slope.

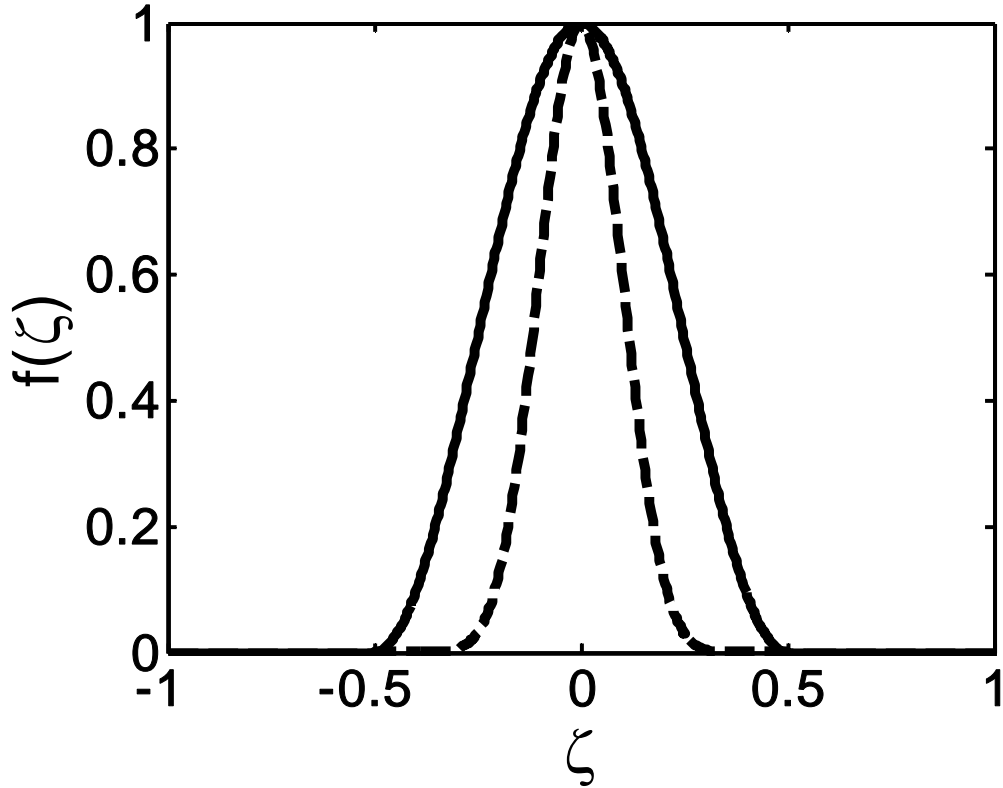

**Fig. 2.** Family of sine power pulses (36): solid line $n = 2$ and dashed line $n = 10$

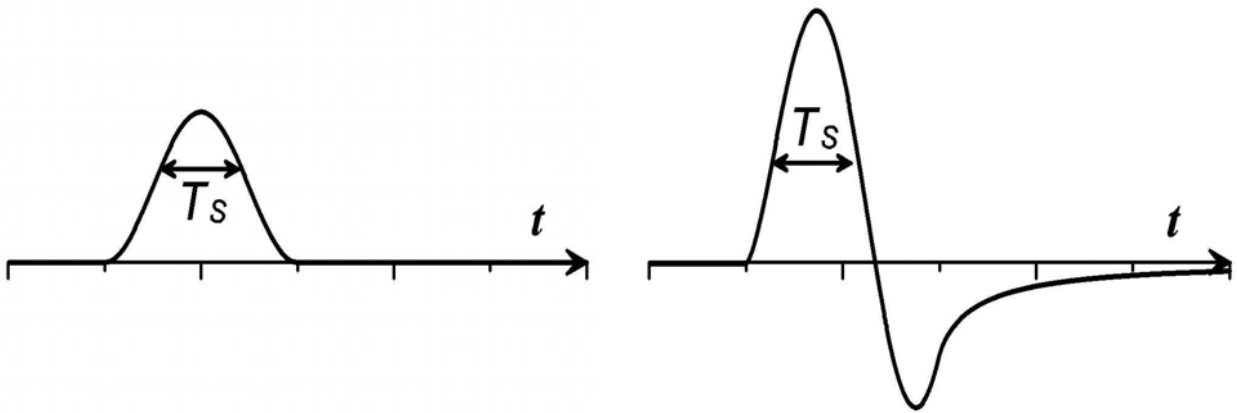

**Fig. 3.** The run-up of symmetric solitary wave on a beach of constant slope. Left panel shows water surface elevation at $x=-L$ right panel shows wave run-up on shore. Notice that the duration Ts of the incident wave, is defined at $x=-L$ and is not necessarily conserved during the run-up and rundown process.

Form factors for run-up $\mu_R^+$ and draw-down $\mu_R^-$ height, run-up $\mu_U^+$ and draw-down $\mu_U^-$ velocity and breaking parameter $\mu_{Br}$, calculated for all sine power pulses (36) with the use of the definition of the characteristic wave length $\lambda_s$ (33) at the 2/3 level of the maximum height, are presented on Fig. 6 in the end of section 4. Calculating means and root-mean-square deviations we obtain the values of form factors for the maximum wave run-up $\mu_R^+ = 3.61 \cdot (1 \pm 0.02)$ and draw-down $\mu_R^- = 1.78 \cdot (1 \pm 0.28)$ that have a fairly limited variation (Table 1) which is given in the end of section 4.

First of all, it is significant that the run-up height is higher than the draw-down height. This feature is observed for all sets of positive impulses. The form factor for the maximum wave run-up in Eq. (35) is almost independent on the power $n$, showing that the influence of the initial wave shape on the extreme run-up characteristics can be made fairly small by an appropriate choice of the characteristic wave length. The above choice of the (significant) wave length reduces the variation of the form factor for the sine power pulses to a remarkably small value, about 2%.

The deepest draw-down is more affected by the wave shape: the relevant form factor varies up to 28%. This feature can be explained by the presence of a complex field of motions in the draw-down phase. A positive wave first executes run-up and only later draw-down (see Fig. 3). Therefore the run-up process is predominantly governed by the incident wave dynamics while the draw-down phenomena occurs under the influence of a set of distributed wave reflections and re-reflections from the slope and consequently it is more sensitive to the wave shape variations.

A similar analysis can be applied to maximum run-up and draw-down velocities of the moving shoreline. Calculated form factors for maximum run-up $\mu_U^+$ and draw-down $\mu_U^-$ velocities are presented on Fig. 7 with triangles. The maximal values for the draw-down velocity are always greater than for the run-up velocity for initial unidirectional impulses. The form factor for the draw-down velocity $\mu_U^- = 6.98 \cdot (1 \pm 0.01)$ is almost constant for all values of $n$ (root-mean-square deviation is 1%) whereas the run-up velocity $\mu_U^+ = 4.65 \cdot (1 \pm 0.30)$ changes in a wider range (±30%); see Table 1.

Variations of the form factor for the breaking parameter are also weak (see Fig. 8, triangles). The case of $n = 2$, corresponding to the discontinuity of the second-order derivative, is excluded since the integral in Eq. (31) diverges. The relevant form factors $\mu_{Br} = 13.37 \cdot (1 \pm 0.10)$ can be considered a constant with a reasonable accuracy (Table 1).

Thus, form factors for the most important parameters such run-up height, draw-down velocity, and to some extent for breaking parameter, are universal and do not depend on the particular shape of a sine power impulse. The variations of form factors for draw-down height and run-up velocity are more significant (about 30%), but they also can be neglected for engineering estimates.

As the second example of the proposed approach we consider the family of solitary waves, described by a following expression

$$f(\zeta) = \operatorname{sech}^n(4\zeta). \qquad n = 1, 2, 3, \ldots \tag{37}$$

These impulses are unlimited in space with exponential decay of the elevation at their ends (see Fig. 4). The case $n = 2$ corresponds to the well-known soliton solution of the Korteweg-de Vries (KdV) equation, which is frequently used as a generic example of shallow water solitary waves.

The run-up of the KdV solitons on a constant beach was studied previously by Synolakis (1987) who presented both experimental and theoretical results. In our notation, the Synolakis formula (Synolakis, 1987) is

$$\frac{R_{max}}{H_0} = 2.8312 \sqrt{\frac{L}{h_0}} \left(\frac{H_0}{h_0}\right)^{1/4}. \tag{38}$$

The "significant" wavelength of the soliton is easily calculated from the well-know analytical expression for a soliton in a constant-depth basin

$$\eta(x) = H_0 \operatorname{sech}^2\left(\sqrt{\frac{3H_0}{4h_0}} \frac{x}{h_0}\right) \tag{39}$$

and has the explicit form:

$$\lambda_s = 4 \operatorname{sech}^{-1}\left(\sqrt{\frac{2}{3}}\right) h_0 \sqrt{\frac{h_0}{3H_0}}, \tag{40}$$

where $\operatorname{sech}^{-1}(z)$ is an inverse function of $\operatorname{sech}(z)$. Substituting the expression for $H_0/h_0$ from (40) into the right-hand side of (38), we obtain

$$\frac{R_{max}}{H_0} = 3.4913 \sqrt{\frac{L}{\lambda_s}}. \tag{41}$$

Our numerical calculations lead to the same value of the form factor, $\mu_R^+ = 3.4913$ at $n = 2$. This example indicates that the theory of soliton run-up on a beach, which leads to a nonlinear relation between the run-up height and the soliton amplitude, is consistent with a general theory of the run-up of solitary waves on a beach and represents a special case.

The form factors for the maximum height of the wave run-up and draw-down for different values of $n$ (Fig. 6) again virtually do not depend on the exponent $n$. This feature suggests that the proposed approach is not sensitive with respect to the shape of the impulses. The form factors, averaged over the range $n = 1 - 20$ are $\mu_R^+ = 3.55 \cdot (1 \pm 0.05)$ for the run-up and $\mu_R^- = 1.56 \cdot (1 \pm 0.28)$ for the draw-down height (Table 1). Notice that these values are close to analogous coefficients for sine power pulses.

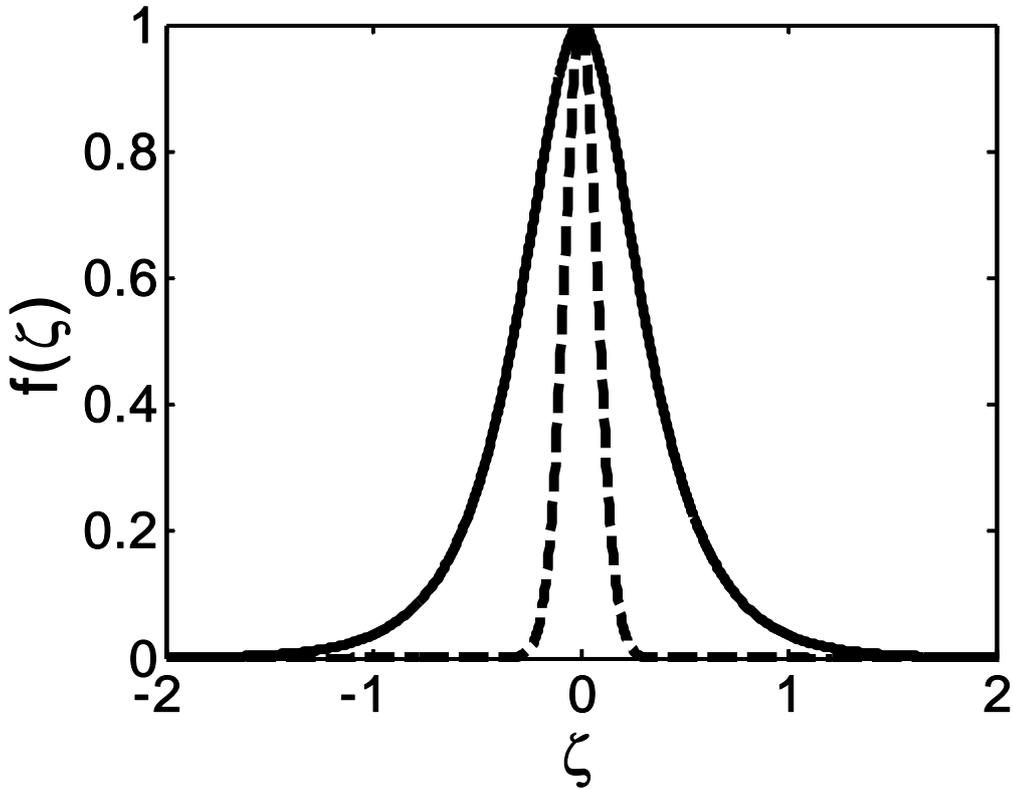

**Fig. 4.** Family of soliton-like impulses (37): solid line $n = 1$ and dashed line $n = 10$

Form factors for run-up and draw-down velocities and breaking parameter (Fig. 7 and 8) are $\mu_U^+ = 4.15 \cdot (1 \pm 0.22)$, $\mu_U^- = 6.98 \cdot (1 \pm 0.02)$, and $\mu_{Br} = 12.90 \cdot (1 \pm 0.03)$ (Table 1). The variation of these parameters for different values of $n$ for soliton-like impulses is to some extent similar to the analogous dependence for sine power pulses. The largest difference is that the run-up velocity form factor for sine-pulses increases with decreasing of the exponent, while the run-up velocity form factor for soliton-like impulses decreases with decreasing of the exponent.

Similar results are obtained for solitary ridges of a Lorentz-like shape with algebraical decay (Fig. 5)

$$f(\zeta) = \frac{1}{\left[1+(4\zeta)^2\right]^n}, \qquad n = 1,2,3,\ldots, \tag{42}$$

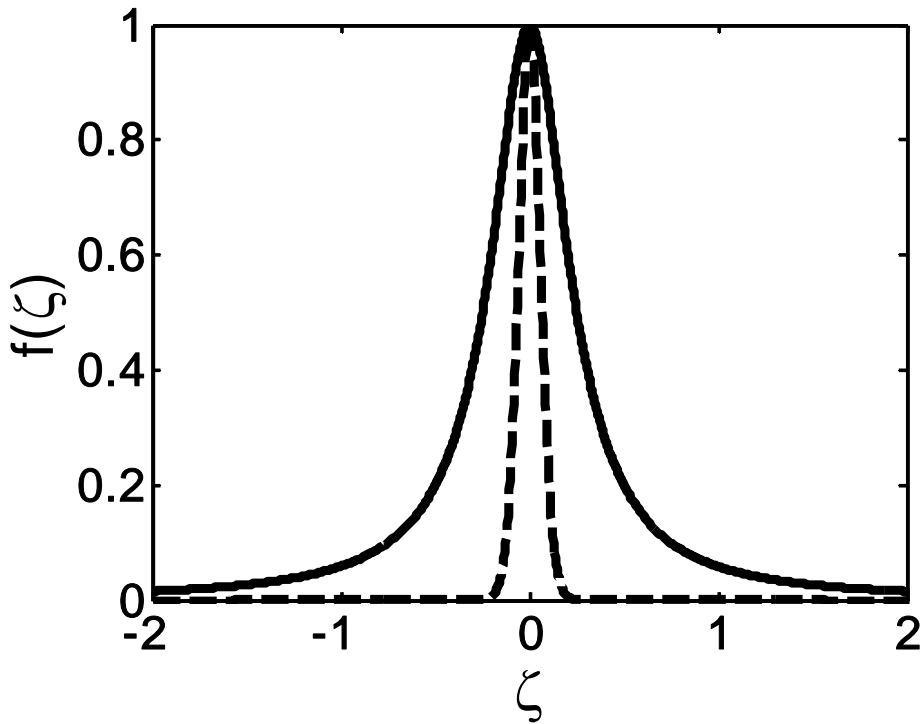

**Fig. 5.** Family of Lorentz-like impulses (42): solid line $n = 1$ and dashed line $n = 10$

**Table 1.** Calculated form factors for different wave shapes

| $\mu$ | Sine power | Soliton power | Lorentz pulse power |
|---|---|---|---|
| $\mu_R^+$ | $3.61 \cdot (1 \pm 0.02)$ | $3.55 \cdot (1 \pm 0.05)$ | $3.53 \cdot (1 \pm 0.08)$ |
| $\mu_R^-$ | $1.78 \cdot (1 \pm 0.28)$ | $1.56 \cdot (1 \pm 0.28)$ | $1.51 \cdot (1 \pm 0.44)$ |
| $\mu_U^+$ | $4.65 \cdot (1 \pm 0.30)$ | $4.15 \cdot (1 \pm 0.22)$ | $4.07 \cdot (1 \pm 0.26)$ |
| $\mu_U^-$ | $6.98 \cdot (1 \pm 0.01)$ | $6.98 \cdot (1 \pm 0.02)$ | $6.99 \cdot (1 \pm 0.04)$ |
| $\mu_{Br}$ | $13.37 \cdot (1 \pm 0.10)$ | $12.90 \cdot (1 \pm 0.03)$ | $12.99 \cdot (1 \pm 0.13)$ |

The calculated run-up and draw-down height form factors for this class of solitary waves (Fig. 6) show some variability in the range of $n = 1 \div 20$. The average values are $\mu_R^+ = 3.53 \cdot (1 \pm 0.08)$ for the run-up and $\mu_R^- = 1.51 \cdot (1 \pm 0.44)$ for the draw-down height (Table 1). The variation of $\mu_R^+$ is still very reasonable. The form factors for run-up and draw-down velocities and breaking parameter (Figs. 7 and 8) show even smaller variation: $\mu_U^+ = 4.07 \cdot (1 \pm 0.26)$, $\mu_U^- = 6.99 \cdot (1 \pm 0.04)$, and $\mu_{Br} = 12.99 \cdot (1 \pm 0.13)$. Their dependence on the exponent $n$ is similar to other families of impulses.

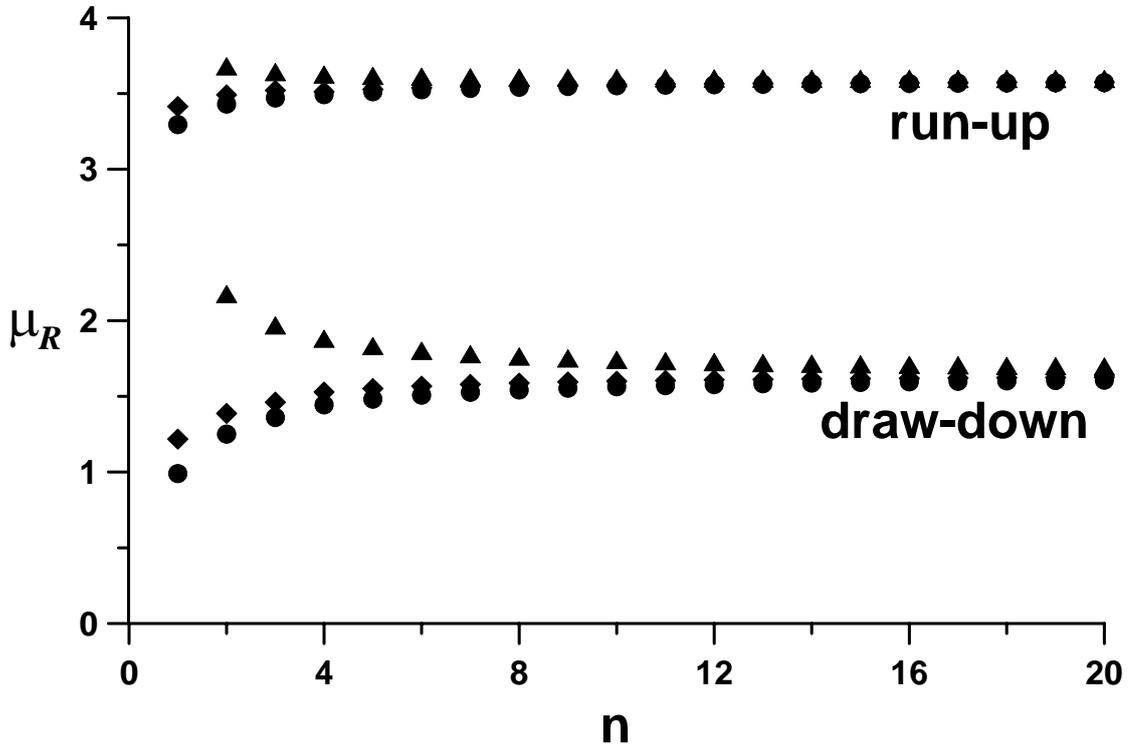

**Fig. 6.** Calculated form factors for the maximum run-up $\mu_R^+$ and draw-down $\mu_R^-$ height for sine power pulses (triangles), soliton-like (diamonds) and Lorentz-like (circles) impulses

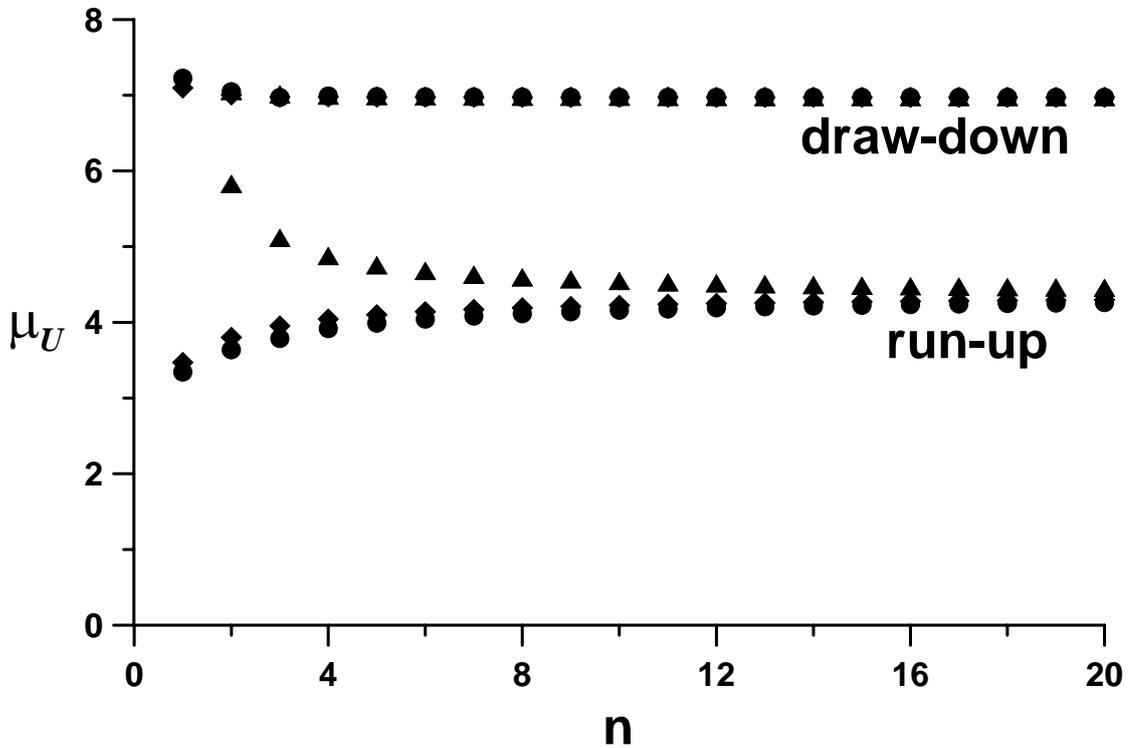

**Fig. 7.** Calculated form factors for the maximum run-up $\mu_U^+$ and draw-down $\mu_U^-$ velocity for sine power pulses (triangles), soliton-like (diamonds) and Lorentz-like (circles) impulses

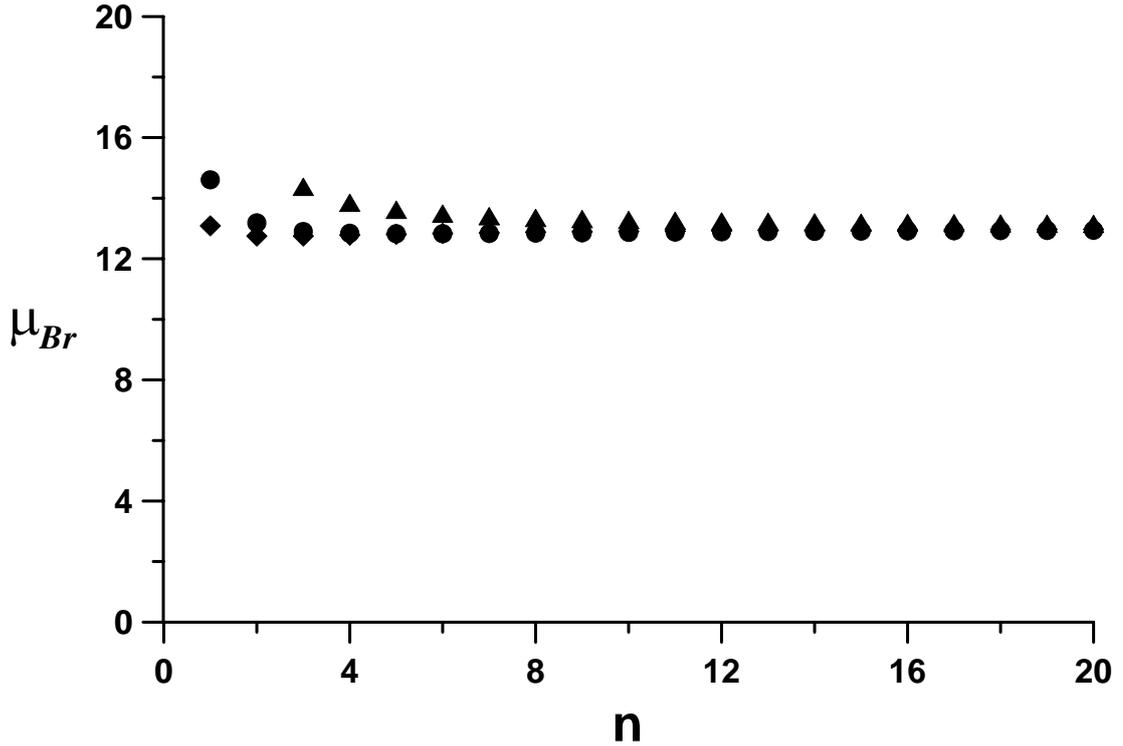

**Fig. 8.** Calculated form factors for the breaking parameter $\mu_{Br}$ for sine power pulses (triangles), soliton-like (diamonds) and Lorentz-like (circles) impulses

## 5. Conclusions

The central outcome from the presented study is that the influence of the initial wave form on maximal run-up characteristics can be almost removed, or made fairly weak, by means of a proper choice of incident wave characteristics. The properties of the features with the largest variation such as the run-up heights, draw-down velocities, and the breaking parameter are at best described with this approach while the other key properties such as the draw-down depths of run-up velocity are reasonably reproduced.

A key result of the study is that the average values of calculated form factors for all concerned classes of symmetrical positive solitary waves (Table 1, Figs. 6-8) and formulas (34) for the maximum run-up and draw-down characteristics of solitary waves virtually do not depend on the form of the incident wave if the wave duration is appropriately defined. This is especially evident for the form factors for the run-up height and draw-down velocity, the variations of which for all the wave classes in question do not exceed 8%.

This property suggests that the definition of the "significant" wave length for solitary waves at the 2/3 level of their maximum height is optimal. In this case the following approximate analogues of formulas (34) for the run-up and draw-down characteristics of the long waves on a beach are universal:

$$R_{\text{run-up}} = 3.5 H_0 \sqrt{\frac{L}{\lambda_s}}, \qquad R_{\text{draw-down}} = 1.5 H_0 \sqrt{\frac{L}{\lambda_s}}, \qquad (43)$$

$$U_{\text{run-up}} = 4.5 \frac{H_0 L}{\lambda_s} \sqrt{\frac{g}{\alpha \lambda_s}}, \qquad U_{\text{draw-down}} = 7 \frac{H_0 L}{\lambda_s} \sqrt{\frac{g}{\alpha \lambda_s}}, \qquad Br = 13 \frac{H_0 L}{\alpha \lambda_s^2} \sqrt{\frac{L}{\lambda_s}}. \qquad (44)$$

Expressions (43-44) can be used for estimates of the run-up and draw-down characteristics of approaching tsunamis as soon as rough estimates for their heights, significant lengths and periods in the open ocean become available.

Finally, we notice that the obtained results hold only for symmetrical solitary waves. If the incident wave has a shape of *N*-wave, also typical for tsunami problem (Tadepalli and Synolakis, 1994, 1996; Tinti and Tonini, 2005), the magnitude of coefficients in "run-up" formulas will differ from given in (43) – (44); see for comparison the difference between run-up heights of soliton and its derivative (Synolakis, 1987; Tadepalli and Synolakis, 1994). If the incident wave is asymmetrical wave with different steepness of the front and back slopes the run-up characteristics depend on the front-slope steepness (Didenkulova et al., 2006, 2007b). Therefore, the universal character of runup characteristics can be achieved "inside" each class of incident wave shapes. As a result, estimates of run-up characteristics require knowledge of only a few "robust" parameters of the incident tsunami wave (positive crest or negative trough, *N*-wave or asymmetrical wave) but not a detailed description of the wave shape. This conclusion is important for practice, as it allows prediction of run-up characteristics of tsunami waves with "unknown" shapes.

**Acknowledgements**
This research is supported particularly by grants from INTAS (06-1000013-9236, 06-1000014-6046), RFBR (08-05-00069, 08-05-72011, 09-05-91222-CT_a), Marie Curie network SEAMOCS (MRTN-CT-2005-019374), Estonian Science Foundation (Grant 5762) and Scientific School of V. Zverev.

## References

**Brocchini, M., and Gentile, R.** (2001) Modelling the run-up of significant wave groups. *Continental Shelf Research,* 21, 1533-1550.

**Carrier, G.F., and Greenspan, H.P.** (1958) Water waves of finite amplitude on a sloping beach. *J. Fluid Mech.*, 4, 97 - 109.

**Carrier, G.F., Wu, T.T., and Yeh, H.** (2003) Tsunami run-up and draw-down on a plane beach. *J. Fluid Mech.*, 475, 79-99.

**Didenkulova, I.I., Zahibo, N, Kurkin, A.A., Levin, B. V., Pelinovsky, E.N., and Soomere, T.** (2006) Run-up of nonlinearly deformed waves on a coast, *Doklady Earth Sciences*, 411, 1241–1243.

**Didenkulova, I.I., Kurkin, A.A., and Pelinovsky, E.N.** (2007a) Run-up of solitary waves on slopes with different profiles. *Izvestiya, Atmospheric and Oceanic Physics*, 43, 384-390.

**Didenkulova, I., Pelinovsky, E., Soomere, T., and Zahibo, N.** (2007b) Run-up of nonlinear asymmetric waves on a plane beach. *Tsunami and Nonlinear Waves* (Ed: Anjan Kundu), Springer, 173-188.

**Didenkulova, I.I. and Pelinovsky, E.N.** (2008) Run up of long waves on a beach: the influence of the incident wave form. *Oceanology,* 48, 1-6.

**Kaistrenko, V.M., Mazova, R.Kh., Pelinovsky, E.N., and Simonov, K.V.** (1991) Analytical theory for tsunami run up on a smooth slope. *Int. J. Tsunami Soc.,* 9, 115 - 127.

**Kânoğlu, U.** (2004) Nonlinear evolution and run-up-rundown of long waves over a sloping beach. *J. Fluid Mech.*, 513, 363-372.

**Kânoğlu, U., and Synolakis, C.** (2006) Initial value problem solution of nonlinear shallow water-wave equations, *Phys. Rev. Letters*, 97 (148501).

**Madsen, P.A., and Fuhrman, D.R.** (2008) Run-up of tsunamis and long waves in terms of surf-similarity. *Coastal Engineering*, 55, 209-223.

**Mazova, R.Kh., Osipenko, N.N., and Pelinovsky, E.N.** (1991) Solitary wave climbing a beach without breaking. *Rozprawy Hydrotechniczne,* 54, 71-80.

**Pedersen, G., and Gjevik, B.** (1983) Run-up of solitary waves. *J. Fluid Mech.*, 142, 283-299.

**Pelinovsky, E., and Mazova, R.** (1992) Exact analytical solutions of nonlinear problems of tsunami wave run-up on slopes with different profiles. *Natural Hazards*, 6, 227 - 249.

**Synolakis, C.E.** (1987) The run-up of solitary waves. *J. Fluid Mech.*, 185, 523-545.

**Synolakis, C.E.** (1991) Tsunami run-up on steep slopes: How good linear theory really is. *Natural Hazards*, 4, 221 – 234.

**Tadepalli, S. and Synolakis, C.E.** (1994) The run-up of N-waves. *Proc. Royal Society London,* A445, 99 - 112.

**Tadepalli, S. and Synolakis, C.E.** (1996) Model for the leading waves of tsunamis. *Physical Review Letters*, 77, 2141 - 2145.

**Tinti, S., and Tonini, R.** (2005) Analytical evolution of tsunamis induced by near-shore earthquakes on a constant-slope ocean. *J. Fluid Mech.*, 535, 33-64.